\documentclass[twocolumn,trackchanges]{aastex62}

\usepackage[latin1]{inputenc}
\usepackage{amsmath}
\usepackage{wasysym}
\usepackage{amsfonts}
\usepackage{amssymb}
\usepackage{graphicx}
\usepackage{listings}
\usepackage{natbib}
\usepackage{comment}
\usepackage{breakurl}
\shorttitle{The Slowly Varying Corona. II}
\shortauthors{Schonfeld et al.}

\begin{document}

\title{The Slowly Varying Corona. II. The Components of F$_{10.7}$ and their Use in EUV Proxies}

\author{S. J. Schonfeld}
\affiliation{NASA Postdoctoral Program Fellow, NASA Goddard Space Flight Center, Heliophysics Science Division, Greenbelt, MD 20771}

\author{S. M. White}
\affiliation{Air Force Research Laboratory, Space Vehicles Directorate, Kirtland AFB, Albuquerque, NM 87117}

\author{C. J. Henney}
\affiliation{Air Force Research Laboratory, Space Vehicles Directorate, Kirtland AFB, Albuquerque, NM 87117}

\author{R. A. Hock-Mysliwiec}
\affiliation{Air Force Research Laboratory, Space Vehicles Directorate, Kirtland AFB, Albuquerque, NM 87117}

\author{R. T. J. McAteer}
\affiliation{New Mexico State University, Department of Astronomy, Las Cruces, NM 88003-8001}

\correspondingauthor{S. J. Schonfeld}
\email{schonfsj@gmail.com}

\begin{abstract}
Using four years of full-disk-integrated coronal differential emission measures calculated in \cite{Schonfeld2017} we investigate the relative contribution of bremsstrahlung and gyroresonance emission in observations of F$_{10.7}$, the 10.7 cm (2.8 GHz) solar microwave spectral flux density and commonly used activity proxy. We determine that the majority of coronal F$_{10.7}$ is produced by the bremsstrahlung mechanism, but the variability observed over individual solar rotations is often driven by gyroresonance sources rotating across the disk. Our analysis suggests that the chromosphere may contribute significantly to F$_{10.7}$ variability and that coronal bremsstrahlung emission accounts for 14.2 $\pm$ 2.1 sfu ($\sim20\%$) of the observed solar minimum level. The bremsstrahlung emission has a power-law relationship to the total F$_{10.7}$ at high activity levels, and this combined with the observed linearity during low activity yields a continuously differentiable piecewise fit for the bremsstrahlung component as a function of F$_{10.7}$. We find that the bremsstrahlung component fit, along with the \ion{Mg}{2} index, correlates better with the observed 5 -- 37 nm spectrum than the common 81 day averaged F$_{10.7}$ proxy. The bremsstrahlung component of F$_{10.7}$ is also well approximated by the moderate-strength photospheric magnetic field parameterization from \cite{Henney2012a}, suggesting that it could be forecast for use in both atmospheric research and operational models.

\end{abstract}


\section{Introduction}
\label{sec:introduction}

Extreme ultraviolet (EUV, 10 -- 121.6 nm) light emitted by the solar atmosphere is absorbed in the terrestrial thermosphere by photoionization, which leads to the creation of the ionosphere as well as the  heating of and increased density in the thermosphere \citep{Tobiska1996a}. Increased incident EUV during periods of increased solar activity can cause communication problems \citep[due to changing radio propagation in the ionosphere,][]{AFHB_Dandekar1985, AFHB_Klobuchar1985, AFHB_McNamara1985} and disrupt satellite operations by increasing satellite surface charging \cite[due to increased ionospheric density,][]{AFHB_Garret1985} and drag \cite[due to increased thermospheric density,][]{DeLafontaine1982}. In addition, the ionosphere forms the boundary layer of the entire terrestrial magnetosphere system that can cause other dynamic terrestrial effects under the influence of space weather events \citep{Schunk2004}. Therefore, regular monitoring of solar EUV emission is of great interest. However, because they are absorbed in the atmosphere, these wavelengths are not observable from the ground, and regular monitoring must be performed by proxy for periods without satellite measurements.

One EUV activity proxy utilized by the solar and terrestrial community is the F$_{10.7}$ index, the $10.7$ cm ($2.8$ GHz) solar microwave spectral flux density. Originally observed by \cite{Covington1947}, F$_{10.7}$ correlates well with solar EUV over month \citep{Chen2012}, year \citep{Balan1993, Lean2011,  Girazian2015, Huang2016}, and solar cycle time scales \citep{Covington1969, Tobiska1991, Bouwer1992, Chen2011a}. F$_{10.7}$ has been a particularly popular observational input to models of the terrestrial upper atmosphere \citep{Bhatnagar1966, Jacchia1971, Hedin1977, Ridley2006, Bowman2008, Bilitza2014} due to its long history and consistent quality \citep{Tapping2013} and remains extensively utilized today even when direct EUV observations are available \citep{Tobiska2008, Bilitza2017}. However, due to observed short-term activity-dependent deviations in F$_{10.7}$ that are not reflected in EUV observations, the microwave observation is typically time-averaged before being used as a model input \citep[e.g.,][]{Hinteregger1981}. One common method is to construct a new smoothed time series
\begin{equation}
\text{F}_{ave} = \left(\text{F}_{10.7} + \text{F}_{81}\right)/2
\label{eqn:F10.7ave}
\end{equation}
where F$_{81}$ is the 81 day centered running average of F$_{10.7}$ \citep{Richards1994}.

One reason for the discrepancy between EUV and F$_{10.7}$ is that while solar EUV emission is generated by collisionally excited atomic emission (with most lines originating in the optically thin corona), there are three distinct emission sources from the non-flaring Sun observed at $10.7$ cm: optically thick (i.e. blackbody) bremsstrahlung emission from the chromosphere \citep{Tapping1987}, optically thin bremsstrahlung emission from the transition region and corona \citep{Landi2003}, and optically thick gyroresonance emission from the corona in the cores of active regions \citep{White1997}. While the chromospheric contribution is thought to be well understood as relating to the solar minimum F$_{10.7}$ level, resolving the relative significance of the bremsstrahlung and gyroresonance components has remained elusive. The lack of resolution is due to conflicting results, with studies relying on imaging analyses typically concluding that bremsstrahlung is the dominant coronal component \citep{Felli1981, Tapping1990, Tapping2003a, Schonfeld2015} while studies utilizing time series analysis conclude that gyroresonance is the dominant mechanism \citep{Schmahl1995, Schmahl1998, DudokdeWit2014a}. Due to the physical processes responsible for the emission, only the bremsstrahlung component of F$_{10.7}$ is related to the EUV emission and directly relevant as an EUV proxy.

Modern solar EUV observations allow the decomposition of F$_{10.7}$  into its components based on the physics of the bremsstrahlung and gyroresonance emission mechanisms. Coronal EUV and microwave bremsstrahlung emission are related to the differential emission measure \citep[DEM,][]{Craig1976}, the plasma density squared integrated over the volume of the optically thin emitting medium as a function of temperature. By determining the DEM using EUV observations, it is possible to calculate the optically thin coronal bremsstrahlung component of F$_{10.7}$ and, with well-constrained assumptions about the chromospheric contribution, predict the gyroresonance component.

In \cite{Schonfeld2017}, hereafter Paper 1, we used the consistent EUV data set provided by the MEGS-A (Multiple Extreme ultraviolet Grating Spectrographs) in the EVE (EUV Variability Experiment) instrument suite on SDO (Solar Dynamics Observatory) to compute a four-year time series of full-disk coronal DEMs. Here, we utilize these DEMs to determine the relative contributions of bremsstrahlung and gyroresonance emission in F$_{10.7}$ during the same period and investigate the implications of these contributions on the use of F$_{10.7}$ as an EUV proxy. We describe the data and processing used to isolate the relevant time series in Section \ref{sec:data}. In Section \ref{sec:f10.7} we describe the procedure used to determine the F$_{10.7}$ emission components and compare these to the F$_{10.7}$ predictions from photospheric magnetic fields presented in \cite{Henney2012a}. We discuss the implications of these findings when using F$_{10.7}$ as an EUV proxy in Section \ref{sec:proxy} and conclude with comments on the continued use of F$_{10.7}$ in Section \ref{sec:conclusion}.

\section{Data}
\label{sec:data}

This investigation relies on a combination of ground and spacecraft data, and careful attention is paid to ensure data consistency and temporal alignment. The data cover just over four years (2010 April 30 to 2014 May 26, described in Section \ref{sec:data:dem}) of the rising phase of solar cycle 24 with a consistent observation time of 2000 UT (described in Section \ref{sec:data:f10.7}). The rare data gaps in the series\footnote{These gaps were caused by CCD bake-out procedures performed on the EVE MEGS-A detectors on 2010 June 16--18, 2010 September 23--27, 2012 March 12--13, 2012 March 19--20.} are filled with a spline interpolation to ensure consistent sampling.


\subsection{The F$_{10.7}$ index}
\label{sec:data:f10.7}

The F$_{10.7}$ index ($10.7$ cm, $2.8$ GHz solar microwave spectral flux density) is observed from the Dominion Radio Astrophysical Observatory near Penticton, Canada. Measurements are made daily at 1700, 2000, and 2300 UT during the summer and 1800, 2000, and 2200 UT during the winter. Each measurement involves a complex observation sequence designed for precision and repeatability and is corrected for (minor) atmospheric absorption. These values are reported in solar flux units (sfu, $10^{-19}\ \text{erg}\ \text{s}^{-1}\ \text{cm}^{-2}\ \text{Hz}^{-1}$) and can be found online at \url{www.spaceweather.gc.ca/solarflux/sx-5-en.php}. A complete description of the F$_{10.7}$ observation and processing procedures is given in \cite{Tapping2013}. For simplicity, we use only the 2000 UT observations except in cases where this observation is missing, in which case we average the other two observations made on the same day to approximate the 2000 UT measurement. We assign the observational error as the greater of 0.5\% \citep{Tapping1994, Tapping2013} and the standard deviation of the three daily measurements. We use the ``adjusted'' values corrected for the Earth's orbital ellipticity, i.e. scaled to the 1 AU mean Earth-Sun separation.

Despite reflecting the combined signal from three different source mechanisms, the F$_{10.7}$ time series has often been treated as the combination of just two phenomenological components, a relatively constant background and a variable contribution related to features associated with magnetic activity such as active regions and plage \citep{Anderson1964, Oster1983a, Oster1983b}. The magnetic activity signal causes the observed rotational modulation in F$_{10.7}$ and is the combination of the bremsstrahlung and gyroresonance components. The constant background is related to the solar minimum level of 66.3 $\pm$ 1.2 sfu that is typically attributed to the chromospheric component. This is not the true minimum F$_{10.7}$ but was calculated by constructing a three-point running median series of F$_{10.7}$ and then averaging the minimum value of this series in each of the six observed solar minima since 1947. It is consistent with methods relating the observed minima in F$_{10.7}$ to the sunspot cycle \citep{Johnson2011, Tapping2011, Bruevich2014}.

\subsection{\ion{Mg}{2} core-to-wing activity index}
\label{sec:data:mgII}

The \ion{Mg}{2} core-to-wing ratio takes advantage of the variation in optical depth across a single broad absorption feature to sample both the chromosphere and photosphere in a single narrow wavelength band. The broad photospheric absorption of the \ion{Mg}{2} h and k lines is relatively constant with the solar cycle while the narrow emission peaks at 2802 \AA\ and 2795 \AA, respectively, are generated in the chromosphere and vary significantly with solar activity \citep{Linsky1970}. Taking the ratio of the intensity in the broad absorption core to the continuum wings provides a measure of solar activity that is insensitive to instrument degradation and calibration effects \citep{Donnelly1994}. \ion{Mg}{2} has been verified as an effective solar activity proxy using spatially resolved imaging \citep{Fredga1971} as well as irradiance observations and has been found to correlate better with certain EUV emission than does F$_{10.7}$ \citep[e.g. for 25 -- 35 nm EUV,][]{Viereck2001}. Here, we use the \ion{Mg}{2} activity proxy (\textbf{version 5, accessed 2019 January 9}) as measured by the second Global Ozone Monitoring Experiment (GOME2) in operation since 2006 \citep{Skupin2005a} and available online at \url{www.iup.uni-bremen.de/gome/solar/MgII_composite.dat}. This data set contains occasional contributions from other instruments to fill in gaps in the primary observation. The data is interpolated to the common 2000 UT sample time.

\subsection{DEMs calculated from EVE MEGS-A spectra}
\label{sec:data:dem}

The EUV Variability Experiment (EVE) Multiple EUV Grating Spectrographs (MEGS)-A observed the solar EUV irradiance spectrum at 5 -- 37 nm aboard the Solar Dynamics Observatory (SDO) satellite \citep{Woods2012} from 2010 April 30 to 2014 May 26 \citep{Pesnell2011}. The availability of these data define the duration of this study. These spectra were collected with a 10 second cadence and have only four data gaps totaling 12 days due to CCD bake-out procedures. The 720 spectra in the two hour window around 2000 UT were used in Paper 1 to compute daily median spectra to remove the effects of short-term variability.

Six Fe emission lines from the median daily spectra were then used to calculate differential emission measures (DEMs, the density squared integrated over the visible coronal volume as a function of temperature) that capture the coronal thermal evolution. These were calculated using the regularized inversion technique described in \cite{Hannah2012} with the CHIANTI 8.0.2 atomic line database \citep{Dere1997, DelZanna2015a}. See Paper 1 Section 3.4 for further details on the DEM calculation procedure and the resulting DEMs in Paper 1 Figure 5. We also determined a conservative error in the DEMs of 15\% which we adopt here as the standard deviation in each temperature bin.

\subsection{F$_{10.7}$ inferred from the photospheric magnetic field}
\label{sec:data:henney}
Because F$_{10.7}$ correlates well with the sunspot number, which is dependent on the photospheric magnetic field configuration, \cite{Henney2012a} used an observation-driven model of the photospheric magnetic field to directly predict F$_{10.7}$. The full-disk-integrated F$_{10.7}$ values were estimated from an empirical model utilizing the sum of the unsigned radial magnetic field magnitude in two different magnetic field strength bins within the Earth-facing regions of global solar magnetic maps \citep{Henney2015}. The maps used for this analysis were generated by the ADAPT \citep[\textbf{A}ir Force \textbf{D}ata \textbf{A}ssimilative \textbf{P}hotospheric Flux \textbf{T}ransport,][]{Arge2013,Hickmann2015} model. ADAPT applies meridional circulation, supergranular diffusion, flux cancellation, and statistical flux emergence in an ensemble of solutions to forward model synoptic magnetic field maps updated continuously using data assimilation. For this study, the ADAPT maps were created using line-of-sight magnetogram data from the Vector Spectromagnetograph \citep[VSM,][]{Henney2009}. The VSM full-disk magnetograms are typically available at a cadence of approximately one per day, and the empirical model predictions were generated using ADAPT maps evolved to 2000 UT to match the F$_{10.7}$, \ion{Mg}{2}, and calculated DEM series.

\section{F$_{10.7}$ emission components}
\label{sec:f10.7}

Essential to separating the F$_{10.7}$ emission components is the (to first order) additive nature of the optically thick chromospheric emission (F$_{chromo}$), the optically thin coronal bremsstrahlung (F$_{brem}$), and the optically thick coronal gyroresonance (F$_{gyro}$):
\begin{equation}
\text{F}_{10.7} = \text{F}_{chromo}+\text{F}_{brem}+\text{F}_{gyro}.
\label{eqn:f10_components}
\end{equation}
The following analysis also relies on the relationship between EUV observations and optically thin coronal microwave bremsstrahlung emission. Their mutual dependence on the coronal DEM and the validated ability to compute the DEM from a set of EUV observations \citep{Guennou2012a, Guennou2012b, Testa2012a} allow the independent determination of F$_{brem}$ when a suitable set of EUV observations exists. Then, a simple assumption about either F$_{chromo}$ or F$_{gyro}$ allows for a complete decomposition of F$_{10.7}$ into its source constituents. The limitations of this methodology are explored in Section \ref{sec:f10.7:radiation}.


\subsection{Calculating the bremsstrahlung component}
\label{sec:f10.7:bremsstrahlung}

\begin{figure*}[t] 
	\centering
	\includegraphics[trim=0cm 0cm 0cm 0cm, scale=1.0]{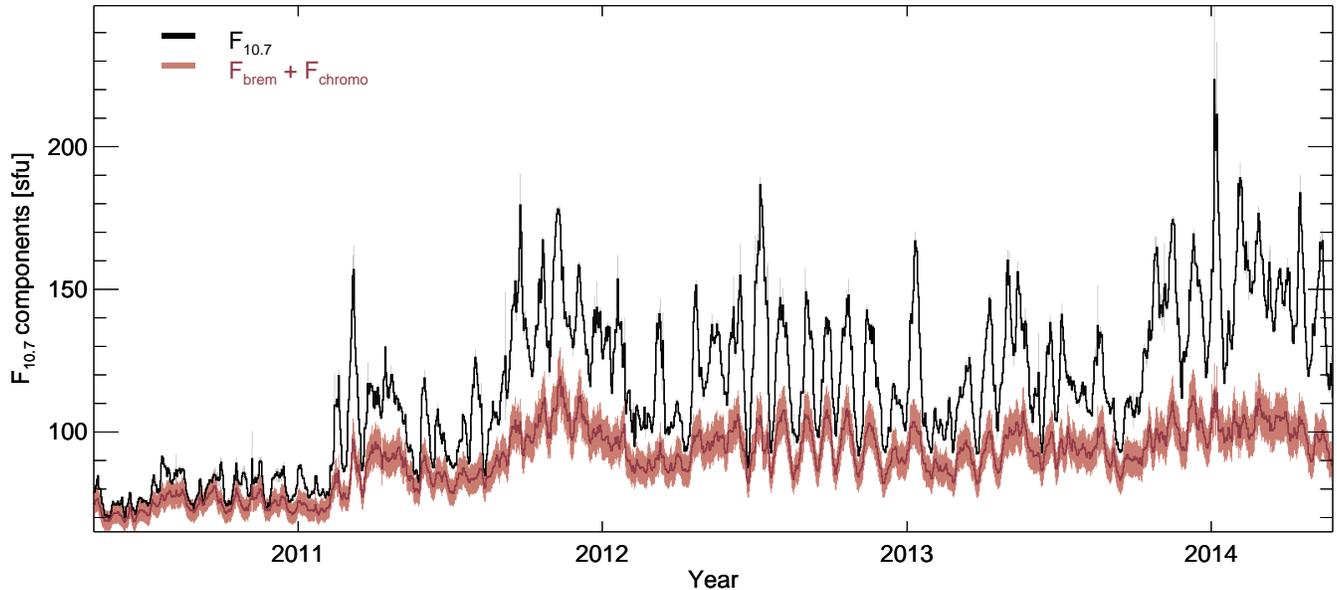}
	\caption{Time series of the adjusted F$_{10.7}$ (black) and the calculated bremsstrahlung component (F$_{brem}$) plus a constant chromospheric component (F$_{chromo}\ =\ 52.1 \pm 2.4$) described in Section \ref{sec:f10.7:linear}. The bremsstrahlung emission accounts for much of the F$_{10.7}$ variability during the solar minimum period (from Paper 1) before February 2011, but a significantly smaller fraction during the solar maximum period after that.}
	\label{fig:f10.7_bremsstrahlung_series}
\end{figure*}

The optically thin coronal bremsstrahlung emission (with units $\text{erg cm}^{-2}\ \text{s}^{-1}\ \text{Hz}^{-1}$) is related to the DEM by
\begin{flalign}
f_{\nu}\ =\ & 9.78\times 10^{-3}\frac{2k_{\text{B}}}{c^2} \left(1+4\frac{n_{\text{He}}}{n_{\text{H}}}\right) \nonumber \\
& \times \int\int T^{-0.5} \text{DEM}(T) G(T)\ \text{d}T\ \text{d}\Omega
\label{eqn:Bremsstrahlung}
\end{flalign}
where $k_{\text{B}}=1.38\times 10^{-16}\ \text{g cm}^{2}\ \text{s}^{-2}\ \text{K}^{-1}$ is Boltzmann's constant, $c=3\times 10^{10}\ \text{cm s}^{-1}$ is the speed of light, $n_{\text{He}}/n_{\text{H}}=0.085$ \citep{Asplund2009} is the density ratio of helium to hydrogen in the emitting medium, $T$ is the temperature in K, $\text{G}(T)=24.5+ln\left(T/\nu\right)$ is the Gaunt factor where $\nu$ is the frequency in Hz, and $d\Omega$ is the solid angle of the source \citep{Dulk1985}. The $\text{DEM}(T)$ (with units $\text{cm}^{-5}\ \text{K}^{-1}$) is
\begin{equation}
\label{eqn:DEM_line}
\text{DEM}(T) = \int\displaylimits_{\text{L}}\frac{d}{dT}\left(n_{\text{e}}n_{\text{H}}\right) d\text{l}
\end{equation}
where L is the optically thin path length of the emitting medium and $n_{\text{e}}$ and $n_{\text{H}}$ are the electron and hydrogen number densities, respectively \citep{Craig1976}. In the case of irradiance observations, equation \ref{eqn:Bremsstrahlung} can be rewritten without the solid angle integral if the DEM is given in the volume integrated form computed in Paper 1 ($\text{cm}^{-3}\ \text{K}^{-1}$):
\begin{equation}
\label{eqn:DEM}
\text{DEM(T)} = \int\displaylimits_{\text{V}}\frac{d}{dT}\left(n_{\text{e}}n_{\text{H}}\right) d\text{V}.
\end{equation}
We calculate the daily optically thin bremsstrahlung emission (F$_{brem}$) for F$_{10.7}$ (at $2.8$ GHz) along with the assumed 15\% error from the DEMs calculated with MEGS-A spectral lines in Paper 1.

F$_{brem}$ plus a constant chromospheric offset (F$_{chromo}$, calculated in Section \ref{sec:f10.7:linear}) is plotted in red in Figure \ref{fig:f10.7_bremsstrahlung_series} and compared with the F$_{10.7}$ in black. While the bremsstrahlung component accounts for much of the F$_{10.7}$ variability during the solar minimum period identified in Paper 1 before February 2011, there is much more variability during solar maximum that cannot be attributed to bremsstrahlung emission. In other words, the F$_{10.7}$ series varies significantly more than the EUV during solar maximum as has previously been noted \citep[by e.g.,][]{Anderson1964}. This has been the primary motivation for using F$_{ave}$ instead of F$_{10.7}$ \citep{Jacchia1964a}.

\begin{figure*}[t] 
	\centering
	\includegraphics*[trim=0cm 1.1cm 0cm 1.5cm, scale=1.0]{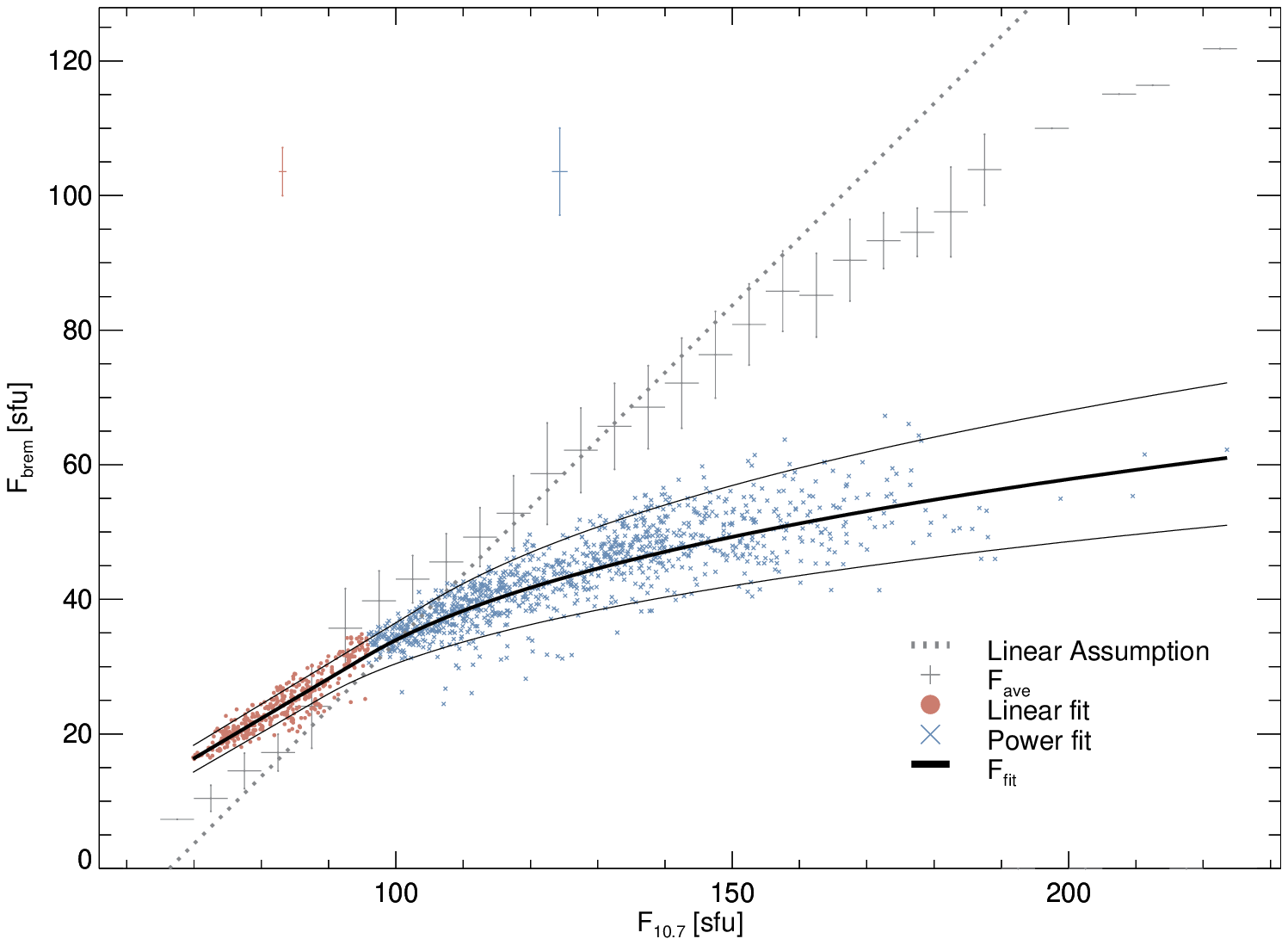}
	\caption{Relationship between the calculated bremsstrahlung emission and the adjusted F$_{10.7}$. The points indicate F$_{brem}$ for the linear (red points) and power (blue \text{\sffamily X}s) regimes of the best-fit trend (black, with $1\sigma$ error trends). The red (blue) cross above the points is the characteristic $1\sigma$ error for the points in the linear (power) portion of the trend. The gray crosses indicate the average F$_{ave}$ in $5$ sfu bins with vertical $1\sigma$ standard deviations. The gray dotted line indicates a unity relationship with F$_{10.7}$ above the solar minimum level, i.e. assuming F$_{10.7}$ is linearly related to EUV. While F$_{ave}$ falls nearly along the unity line, F$_{brem}$ deviates significantly.} 
	\label{fig:f10.7_bremsstrahlung_correlation}
\end{figure*}                                                                                                                                                                             

F$_{brem}$ is plotted as the red points and blue crosses versus F$_{10.7}$ in Figure \ref{fig:f10.7_bremsstrahlung_correlation}. It is immediately obvious that this relationship is well constrained and linear at low activity levels but becomes more variable and nonlinear at high activity levels. We fit this bremsstrahlung distribution with a continuously differentiable piecewise function that is linear at low activity and a power function at high activity such that
\begin{equation}
\text{F}_\text{fit}=\begin{cases}
	a+b\text{F}_{10.7}, & \text{for } \text{F}_{10.7} < c\\
	d\left(\text{F}_{10.7}-e\right)^{f}, & \text{for } \text{F}_{10.7} \ge c
	\end{cases}
\label{eqn:piecewise}
\end{equation}
is the best-fit bremsstrahlung component of F$_{10.7}$. The best least-squares solution has $a=-25.1\pm 1.1$, $b=0.592\pm 0.013$, $c\approx96$, $d=13.9\pm 1.2$, $e\approx80$, and $f=0.298\pm 0.019$ and is found using the IDL MPFIT package of \cite{Markwardt2009}. Requiring continuity and smoothness (continuous first derivative) in this functional form provides two constraints to the fit and the function has been arranged so that $c$ and $e$ are uniquely determined by the other best-fit parameters and do not have derived errors. In Figure \ref{fig:f10.7_bremsstrahlung_correlation}, the red points are fit by the linear component, the blue \text{\sffamily X}s are fit by the power component, and the solid black line indicates the best-fit function with $1\sigma$ errors. F$_{brem}$ is very different from both F$_{ave}$ (indicated by the gray crosses that show $5$ sfu binned averages with $1\sigma$ standard deviations) and the canonically assumed linear EUV--F$_{10.7}$ relationship (dotted gray line).

\subsubsection{Bremsstrahlung and photospheric magnetic fields}
\label{sec:f10.7:bremsstrahlung:henney}

\begin{figure}[t] 
	\centering
	\includegraphics*[trim=0cm 0cm 0cm 0cm, scale=1.0]{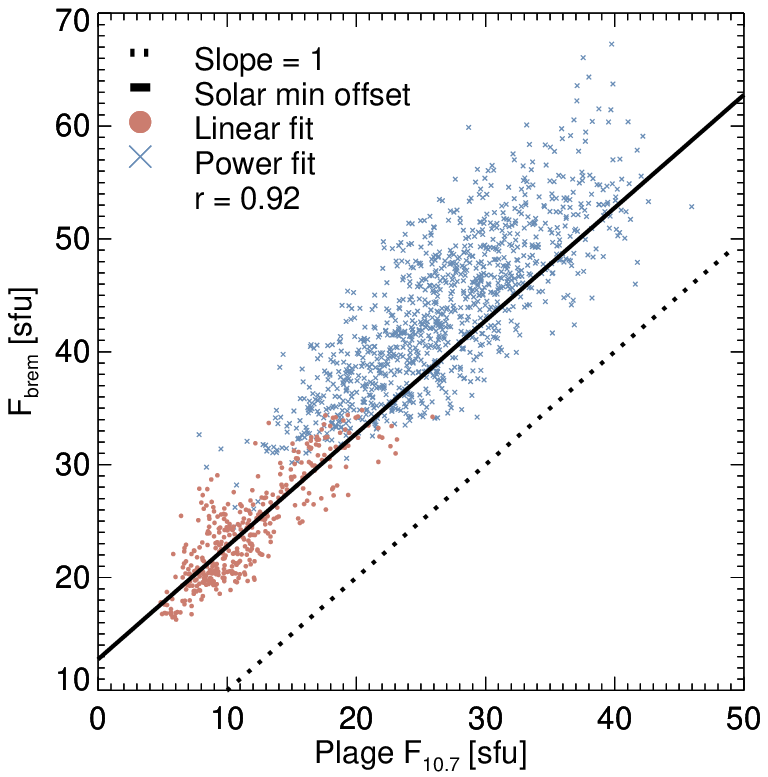}
	\caption{Correlation between F$_{brem}$ (with the same colors and symbols as Figure \ref{fig:f10.7_bremsstrahlung_correlation}) and the predicted F$_{10.7}$ from plage region photospheric magnetic fields (defined as radial magnetic fields between 20 and 150 G) following the method of \cite{Henney2012a}. The dotted black line indicates equality between the two parameters while the solid black line indicates that same equality after accounting for a solar minimum offset including the solar minimum bremsstrahlung emission calculated in Section \ref{sec:f10.7:linear}.}
	\label{fig:henney_bremsstrahlung}
\end{figure}                                                                                                                                                                            

The procedure of \cite{Henney2012a} involved parameterizing F$_{10.7}$ using observations of the photospheric magnetic field magnitude. This was done using two components, a "plage" component with fields between 25 and 150 G, and an "active region" component with fields greater than 150 G. The moderate magnetic field strength of the "plage" magnetic fields are associated with increased coronal plasma density \citep[e.g.,][]{Kelch1978} and it is therefore natural to expect correlation with F$_{brem}$. This relationship is plotted in Figure \ref{fig:henney_bremsstrahlung} and shows a generally linear correlation with a Pearson coefficient of 0.92 and a slope near one, but with significantly more F$_{brem}$ than F$_{10.7}$ plage-field component. This is due to the assumption in \cite{Henney2012a} that attributed all solar minimum F$_{10.7}$ to a constant (presumably chromspheric) component whereas we find a solar minimum coronal bremsstrahlung contribution (details in Section \ref{sec:f10.7:linear}). Correcting for this difference in the solar minimum level yields much better agreement, indicated by the solid black line (compared to the dotted line without this correction) that represents equality between F$_{brem}$ and the F$_{10.7}$ plage field component.

This agreement suggests that the method of \cite{Henney2012a} naturally identifies F$_{brem}$ based on the observed photospheric magnetic field strength. This is particularly valuable because of the demonstrated ability to evolve ADAPT photospheric magnetic field maps to predict F$_{10.7}$ \citep{Henney2012a} and EUV irradiance \citep{Henney2015}. The correlation in Figure \ref{fig:henney_bremsstrahlung} indicates it is possible to approximate F$_{brem}$ from photospheric magnetic field measurements even when EUV observations are not available.

\subsection{Utilizing the low-activity linearity}
\label{sec:f10.7:linear}

\begin{figure*}[t] 
	\centering
	\includegraphics*[trim=0cm 0cm 0cm 0cm, scale=1.0]{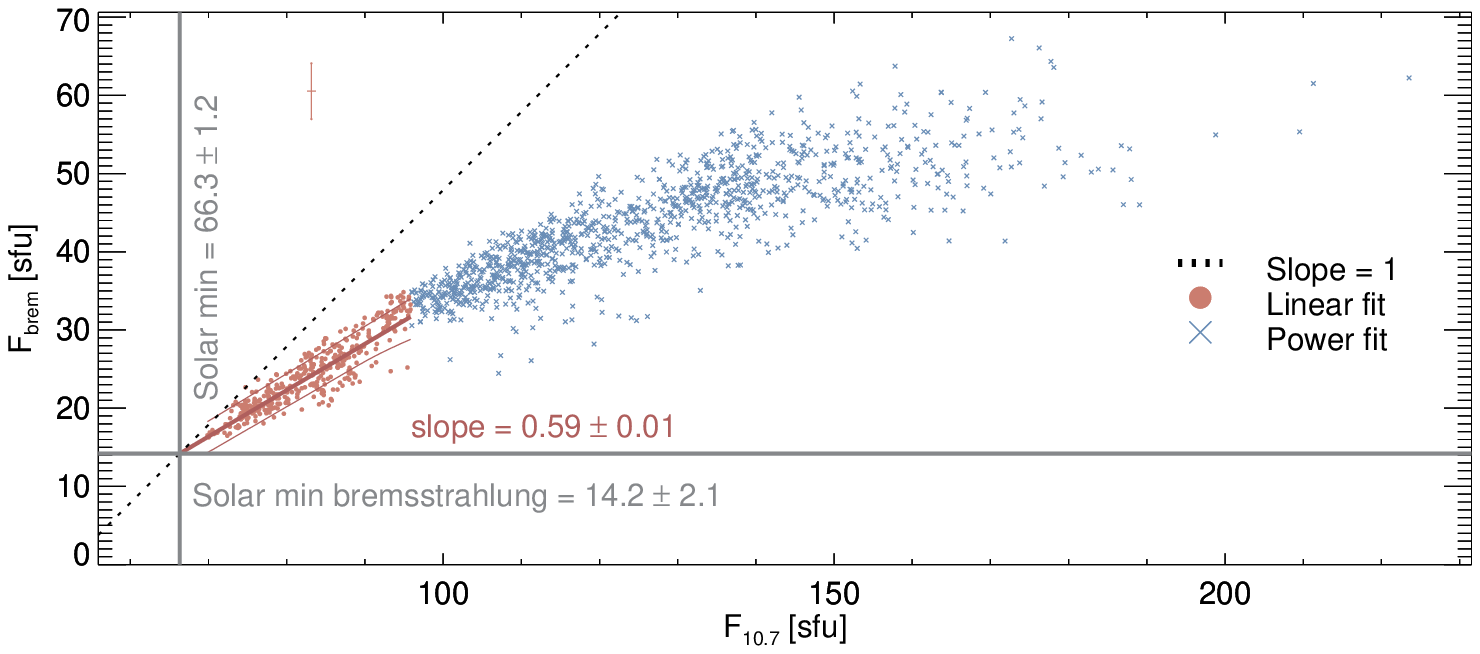}
	\caption{Linear (power) relationship between F$_{brem}$ and adjusted F$_{10.7}$ during the low (high) activity periods plotted as red points (blue \text{\sffamily X}s). The red cross is the characteristic $1\sigma$ error for the points in the linear regime. The vertical gray line indicates the solar minimum F$_{10.7}$ level and the horizontal gray line marks the solar minimum coronal bremsstrahlung contribution calculated by extrapolating the linear fit to the solar minimum level outside the observation range. The slope of this linear fit is significantly less than one, indicating that F$_{brem}$ does not account for all of the F$_{10.7}$ variation observed in the linear regime.}
	\label{fig:f10_linear}
\end{figure*}                                                                                                                                                                             

There is significant value in examining in detail the linear component of the bremsstrahlung--F$_{10.7}$ relationship plotted again as red points in Figure \ref{fig:f10_linear}. First, by extrapolating this linear trend back to the $66.3 \pm 1.2$ sfu F$_{10.7}$ solar minimum level, we find that $14.2 \pm 2.1$ sfu of it is due to coronal bremsstrahlung emission, implying $52.1 \pm 2.4$ sfu is a constant chromospheric component. This is important because time series analyses typically consider only the rotationally variable component of F$_{10.7}$ \citep{Schmahl1995, Schmahl1998, DudokdeWit2014a}, but this necessarily excludes the solar minimum bremsstrahlung component that indicates significant coronal EUV emission, even at solar minimum. In addition, the slope of the relationship of this linear regime is informative for the nature of F$_{10.7}$ emission during the solar minimum period. If bremsstrahlung emission accounted for all the variation in F$_{10.7}$ observed during this period, the slope of this relationship would be unity. The fact that it is less than one has three potential explanations: gyroresonance emission contributed a significant fraction of F$_{10.7}$ even during the low-activity period, the iron abundance used to calculate the DEMs in Paper 1 was too large, or the contribution from the chromosphere is also activity dependent. Each of these explanations are explored in the following subsections.

\subsubsection{Magnetic field dependence of gyroresonance emission}
\label{sec:f10.7:linear:gyroresonance}
Due to the nature of the gyroresonance mechanism, coronal observations at a fixed frequency only detect gyroresonance emission from discrete, narrow layers in the atmosphere. In particular, emission is only observed at harmonics of the gyrofrequency (in MHz) given by:
\begin{equation}
\nu_{B} = 2.80B
\label{eqn:gyrofrequency}
\end{equation}
where B is the total magnetic field strength in G \citep{White1997}. The fundamental emission at a gyrofrequency of 2.8 GHz (F$_{10.7}$) is produced in magnetic fields of 1000 G that only exist at altitudes below the optically thick floor. Consequently, emission is typically observed from the second, third and fourth harmonics which, for 2.8 GHz observations, originate in magnetic field layers with B = 500, 333, and 250 G, respectively \citep{White2005a}. In addition, due to the decrease of the coronal magnetic field with altitude \citep{Kuridze2019}, it is reasonable to expect photospheric magnetic fields of at least 500 G in order to yield appreciable coronal gyroresonance emission at 2.8 GHz.

\begin{figure}[t] 
	\centering
	\includegraphics*[trim=0cm 0cm 0cm 0cm, scale=1.1]{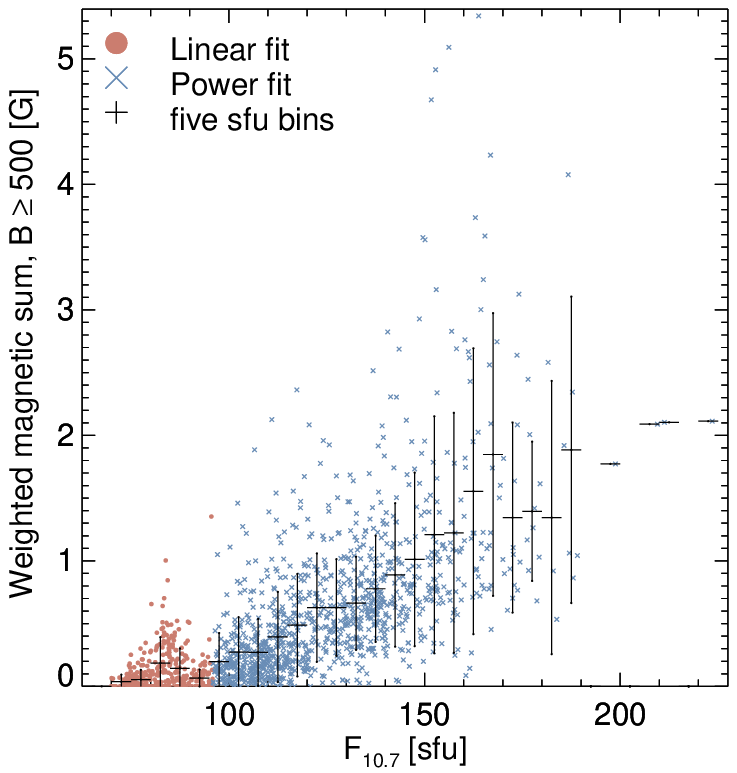}
	\caption{Correlation between the area weighted sum of photospheric magnetic fields greater than 500 G and adjusted F$_{10.7}$. The red points and blue \text{\sffamily X}s indicate the same activity split as Figure \ref{fig:f10.7_bremsstrahlung_correlation} and the black crosses are the 5 sfu binned values with the associated standard deviation within the bin. These indicate that the magnetic fields necessary to produce gyroresonance emission are at a minimum during the low-activity linear regime and, within each 5 sfu bin in this regime, the average magnetic field sum is within a standard deviation of zero, despite being strictly positive.}
	\label{fig:henney_500}	
\end{figure}

Figure \ref{fig:henney_500} shows the area weighted sum of photospheric magnetic fields greater than 500 G, which we interpret as the total magnetic field capable of producing gyroresonance emission, plotted against F$_{10.7}$. The black crosses indicate the average over 5 sfu bins with the vertical error bars representing the 1$\sigma$ standard deviation in the measured values. From this we can see a significant increase in the 500 G magnetic field sum during the power-fit period and that the average during the linear-fit regime is within a standard deviation of zero. This suggests that while isolated low activity days may have gyroresonance emission, the majority of the linear-fit period should have essentially no gyroresonance emission component in F$_{10.7}$. This is consistent with an intuitive understanding that during solar minimum there are relatively few active regions with intense magnetic fields, and long periods with no visible active regions at all.

\subsubsection{The effect of coronal iron abundance}
\label{sec:f10.7:linear:abundance}
Another possible explanation for the non-unity slope in Figure \ref{fig:f10_linear} is a true iron abundance significantly different from $N_{\text{Fe}}/N_{\text{H}} = 1.26\times10^{-4}$ used in Paper 1, an enhancement by about a factor of four above photospheric values \citep{Feldman1992}. This is a commonly used value for the ``coronal'' iron abundance \citep[e.g.,][]{Kashyap1998, White1999, Landi2003, Landi2008, Warren2012, Schonfeld2015}, but other work has suggested the coronal enhancement may be less, as low as a factor of two \citep[e.g.,][]{Zhang2001, Warren2009, DelZanna2013b, Guennou2013} or even no enhancement in coronal holes \citep{ChiuderiDrago1999}. The DEMs in Paper 1 were calculated using only iron emission lines and are thus inversely related to the coronal iron abundance (see Paper 1 Section 3.3 for more details). This means the calculated bremsstrahlung emission which is related to the total plasma density (primarily hydrogen) also scales inversely with the iron abundance. It is therefore possible to ``recalculate'' the bremsstrahlung prediction with different coronal iron abundances by applying a simple multiplicative scaling.


\begin{figure*}[t] 
	\centering
	\includegraphics*[trim=0cm 0cm 0cm 0cm, scale=1.0]{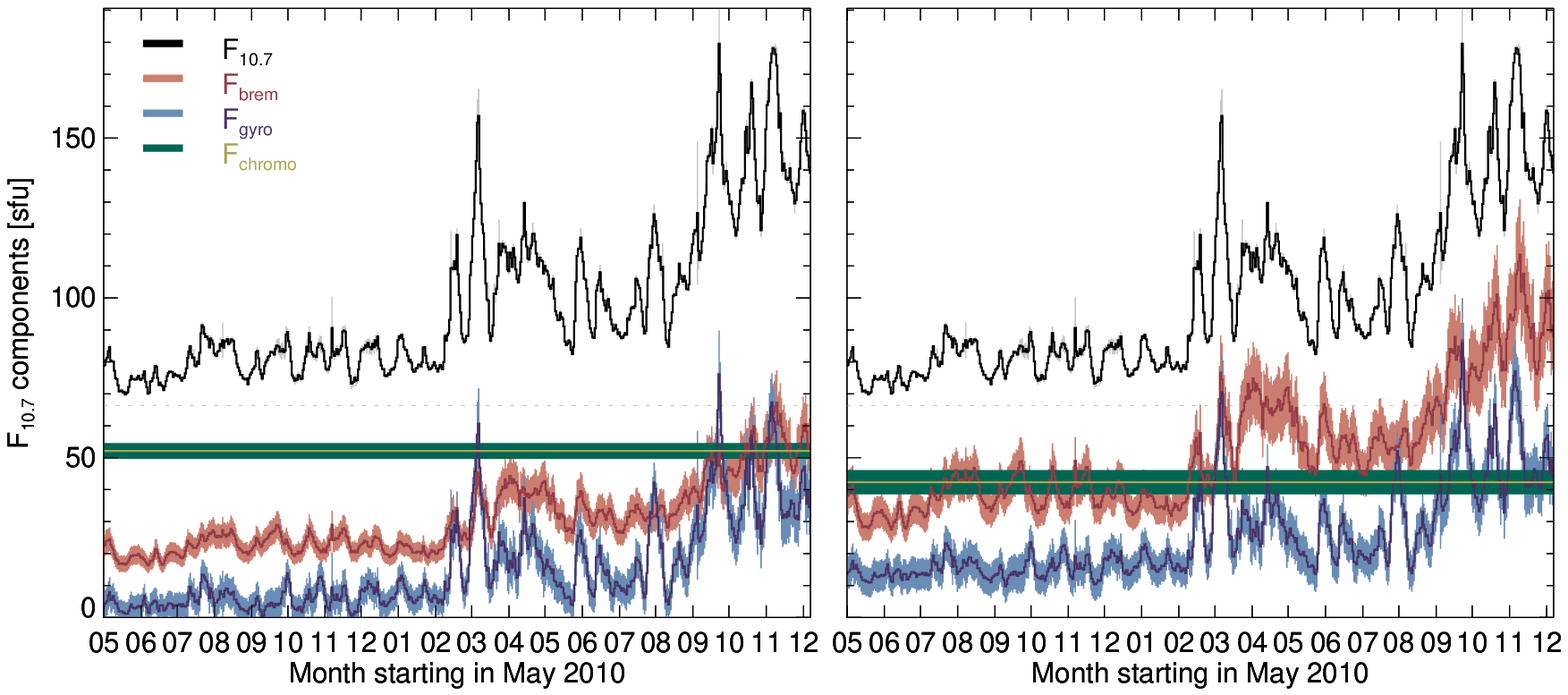}
	\caption{Time series of adjusted F$_{10.7}$ and the calculated bremsstrahlung, gyroresonance, and chromospheric components for \text{left}: a standard coronal iron abundance 3.89 times the photospheric level \citep{Feldman1992} and \textit{right}: an iron abundance 2.3 times the photospheric level as calculated in \ref{sec:f10.7:linear:abundance}. These figures demonstrate that calculating F$_{brem}$ with a decreased coronal iron abundance leads to increased F$_{gyro}$, in conflict with the implications of \ref{sec:f10.7:linear:gyroresonance}.}
	\label{fig:f10_comp_abundances}
\end{figure*}                                                                                                                                                                             

Correcting the slope in Figure \ref{fig:f10_linear} to unity by scaling F$_{brem}$ corresponds to a coronal iron abundance of $N_{\text{Fe}}/N_{\text{H}} = 7.45\times10^{-5}$, a factor of 2.3 enhancement above the photosphere. In addition to modifying the slope of the bremsstrahlung-F$_{10.7}$ relationship, this also changes the expected solar minimum bremsstrahlung contribution to $24.0 \pm 3.6$ sfu, suggesting only $42.3 \pm 3.8$ sfu of chromospheric emission. Considering equation \ref{eqn:f10_components}, we use F$_{10.7}$ and the known bremsstrahlung and chromosphere components to calculate the expected gyroresonance emission. F$_{10.7}$ and its three components are plotted in Figure \ref{fig:f10_comp_abundances} for the standard coronal abundance (left panel) and the modified coronal abundance (right panel). These figures demonstrate that not only does decreasing the iron abundance lead to significantly more bremsstrahlung emission as expected, it also suggests significantly more gyroresonance emission due to reduced chromospheric emission.

This greater-than-10-sfu minimum gyroresonance contribution is inconsistent with our earlier analysis on a number of levels. First, it is at odds with the physical nature of gyroresonance emission, which is concentrated in active regions \citep{Schonfeld2015} that are not consistently present on the visible solar disk during solar minimum. It is also in conflict with the suggestion from Figure \ref{fig:henney_500} that there are insufficient strong magnetic fields to produce significant gyroresonance emission during the low activity period. Finally, the assumption behind forcing the linear fit component of the bremsstrahlung series into agreement with F$_{10.7}$ was that there should be no gyroresonance emission during this period, which is wholly opposite to the effect created by this correction. The conflict of this significant redistribution of the emission components with these other observable characteristics suggests that decreasing the coronal iron abundance to force a unity bremsstrahlung-F$_{10.7}$ relationship during solar minimum is inappropriate.

\subsubsection{The implications of chromospheric variability}
\label{sec:f10.7:linear:chromosphere}

\begin{figure}[t] 
	\centering
	\includegraphics*[trim=0cm 0cm 0cm 0cm, scale=1.1]{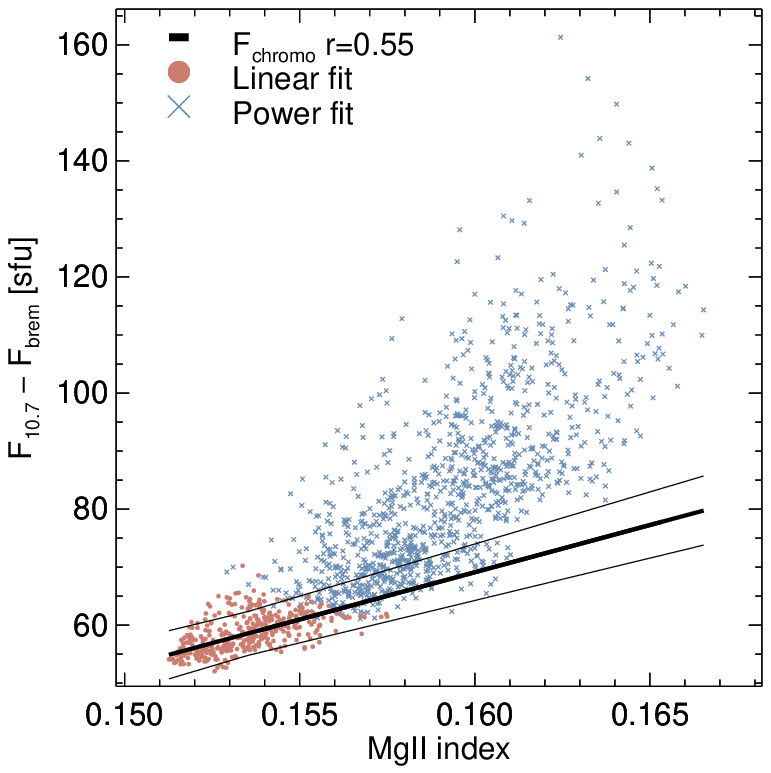}
	\caption{Relationship between the non-bremsstrahlung F$_{10.7}$ and the \ion{Mg}{2} index. The linear-fit component of this remaining emission (which is assumed to be produced purely by optically thick bremsstrahlung in the chromosphere) is plotted as red points, and the power component with gyroresonance contributions is plotted as blue \text{\sffamily X}s. The black lines indicate the best-fit correlation of the linear component (and the associated $1\sigma$ errors), which is used to calculate the variable chromosphere in Figure \ref{fig:component_series_variable_chromosphere}.}
	\label{fig:MgII_correlation}
\end{figure}

The final explanation of the less than unity slope in Figure \ref{fig:f10_linear} is the variability of the chromosphere. In all previous analyses, we have assumed that the chromosphere is constant, but this is obviously suspect since the chromosphere is observed to be highly dynamic \citep[][and references therein]{Hall2008} and the source of the \ion{Mg}{2} $280$ nm doublet that is itself used as a solar activity proxy \citep{Heath1986}. We can use the linear regime of Figure \ref{fig:f10_linear} assuming only contributions from the chromosphere and coronal bremsstrahlung to construct an estimated variable chromosphere by finding the best-fit linear correlation between $\text{F}_{chromo}=\text{F}_{10.7}-\text{F}_{brem}$ and the \ion{Mg}{2} index. This is shown in Figure \ref{fig:MgII_correlation} which reveals a roughly linear relationship with Pearson correlation coefficient $\text{r}=0.55$ in the linear bremsstrahlung regime. Because this is a small correction relative to the bremsstrahlung variability, this relatively low correlation should be sufficiently accurate given the $15\%$ errors inherited from the DEM calculation. Using the observed \ion{Mg}{2} during these four years, this correlation allows us to estimate the contribution from a variable chromosphere even during solar maximum when F$_{gyro}$ is significant.

\begin{figure*}[t] 
	\centering
	\includegraphics*[trim=0cm 0cm 0cm 0cm, scale=1.0]{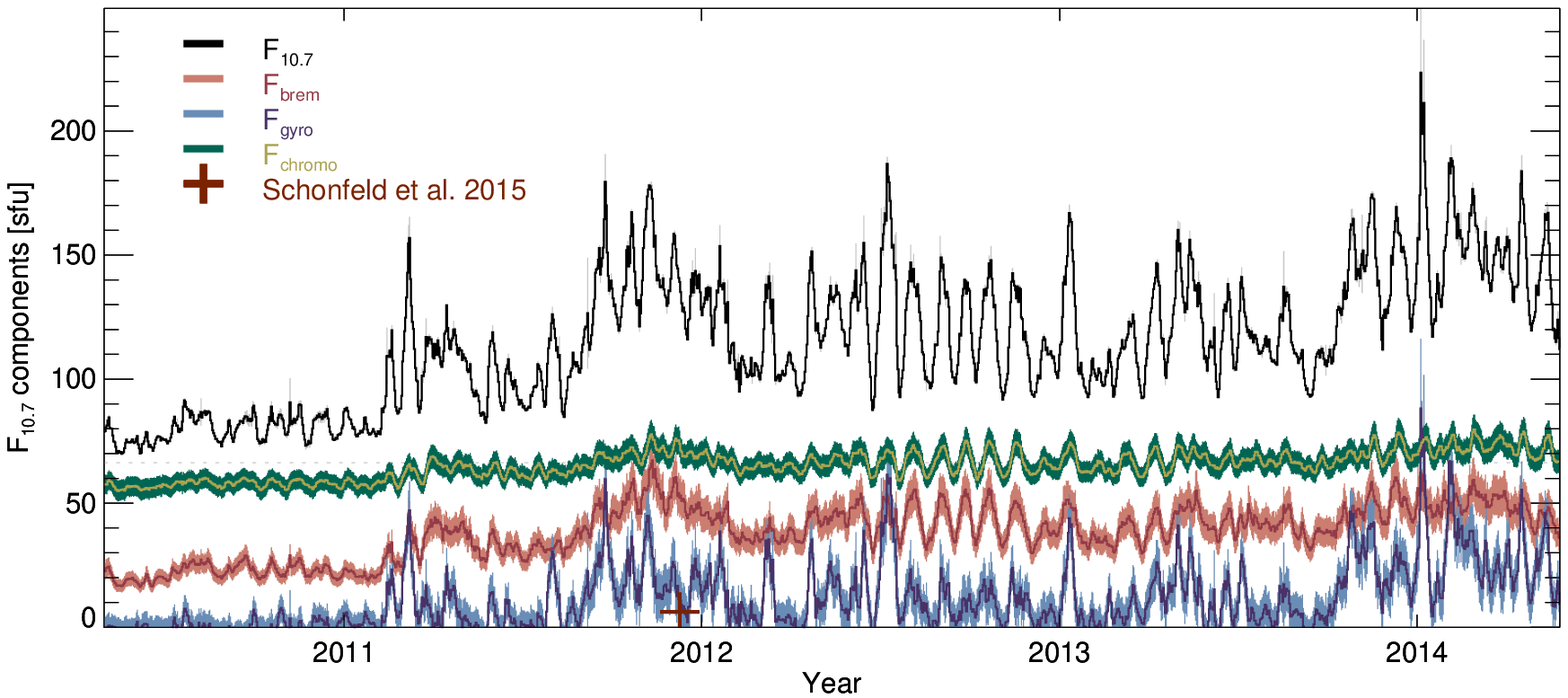}
	\caption{Similar to Figure \ref{fig:f10.7_bremsstrahlung_series} but with a variable chromsphere calculated using the correlation in Figure \ref{fig:MgII_correlation} and the corresponding calculated gyroresonance component. Utilizing a variable chromosphere results in essentially no F$_{gyro}$ during solar minimum and many rotational minima during solar maximum. The maroon plus symbol indicates the $6.2 \pm 0.3$ sfu of gyroresonance emission measured by the VLA on 2011 December 9 \citep{Schonfeld2015}, which is consistent with the F$_{gyro}$ value within the uncertainties.}
	\label{fig:component_series_variable_chromosphere}
\end{figure*}

The F$_{10.7}$ time series with components determined using this best-fit variable chromosphere is shown in Figure \ref{fig:component_series_variable_chromosphere}. Including this variable chromosphere increases its contribution compared to the assumed constant chromosphere, leading to a decrease in the calculated gyroresonance component. During solar maximum, this accentuates the effect of individual solar rotations, with these series suggesting that many solar rotations have no gyroresonance during their local minima. This variable chromosphere also reduces the gyroresonance contribution during the solar minimum period (as identified in Paper 1 before 2011 February 8 when solar activity abruptly turns on) to a level that is consistent with no gyroresonance within the uncertainties. This is by construction, since we assumed no gyroresonance in the linear regime when calculating the variable chromosphere.

This variable chromospheric correction also brings the calculated gyroresonance component into better agreement with a Very Large Array (VLA) full-disk image at 2.782 GHz from 2011 December 9 \citep{Schonfeld2015}. That analysis found $6.2 \pm 0.3$ sfu of gyroresonance emission and is delineated as the maroon cross in Figure \ref{fig:component_series_variable_chromosphere}. When using a variable chromosphere, this disk-integrated time series analysis suggests $16.2 \pm 9.9$ sfu of gyroresonance, compared to $34.9 \pm 8.3$ sfu with a constant chromosphere. The reduced gyroresonance calculated with the variable chromosphere is consistent (given the uncertainties) with the previous imaging analysis.

These results reconcile the apparent contradiction between previous analyses that disagreed about the relative contribution of bremsstrahlung and gyroresonance emission in F$_{10.7}$. It suggests that those studies that attribute the majority of coronal F$_{10.7}$ to bremsstrahlung \citep{Felli1981, Tapping1990, Tapping2003a} may be correct during all but the most active periods. Furthermore, the studies that suggest gyroresonance dominates the F$_{10.7}$ variability \citep{Schmahl1995, Schmahl1998, DudokdeWit2014a} may also be correct, particularly during active periods when the variation over a single rotation due to gyroresonance emission can be greater than the total bremsstrahlung. The cause of these previous disagreements is the low level background coronal bremsstrahlung which persists throughout the solar cycle. This contribution is apparent in images but was subtracted during previous time series analyses of F$_{10.7}$ variability.


\subsection{Complicating details of optical depth}
\label{sec:f10.7:radiation}

The analysis outlined in this paper relies on the assumption that the coronal volume observed in the EUV is the same as is observed by F$_{10.7}$. It is from this assumption that we developed equation \ref{eqn:f10_components}. However, there are two optical depth effects that violate this assumption, both leading to more plasma being visible in the EUV than at F$_{10.7}$.

First, the chromosphere becomes optically thick at higher altitudes in the microwave \citep[e.g.,][]{Gary1996, Selhorst2005} than it does for the EUV lines used to calculate the DEMs in Paper 1. Because there is no EUV emission in the chromosphere from the log(T[K]) $>$ 5.5 lines used to compute the DEMs, this has no effect on the observed emission from the solar disk. However, this also means that the disk of the Sun appears larger at F$_{10.7}$ and therefore more plasma is visible in the EUV behind the solar limb. EUV sources become visible sooner and remain visible longer when rotating onto and off of the solar disk, respectively.  Based on the rate of solar rotation and the $\approx$30\arcsec\ increased limb altitude at F$_{10.7}$ \citep{Furst1979}, the duration of this effect for a given source is typically less than a day and depends on the source altitude and latitude. Since this analysis is performed with a daily cadence, the magnitude of this effect near the limb is highly dependent on the level of activity in a narrow region just beyond the limb. \cite{Schonfeld2015} found that during an active period with many active regions on the limb, F$_{10.7}$ was depressed by $\sim6\%$ due to the increased size of the solar disk compared to the EUV.

Second, gyroresonance becomes optically thick in the corona. This means that a gyroresonance source in the corona blocks emission from the underlying chromosphere and whatever coronal bremsstrahlung emission occurs behind it from the observer's perspective. At the same time, EUV light from these lower layers reaches the observer uninhibited. Gyroresonance emission therefore also reduces the coronal volume probed by F$_{10.7}$. These optically thick gyroresonance layers always have a greater surface brightness than the underlying chromosphere and coronal bremsstrahlung and therefore blocking the lower layers actually causes less of an increase in the disk-integrated F$_{10.7}$ than if these components were truly additive. Fortunately for this analysis, gyroresonance sources are relatively small and therefore this is expected to have only a small effect on the measured F$_{10.7}$. 

\section{F$_{10.7}$ as an EUV proxy}
\label{sec:proxy}
Numerous studies \citep[e.g.,][]{Tapping1987, Tobiska1991, Balan1993, Tobiska2001, Chen2012, Bruevich2014, Huang2016} have previously identified the imperfect relationship between EUV and F$_{10.7}$ and corrective strategies (such as the utilization of F$_{ave}$) have been developed to adapt F$_{10.7}$ for use as an input to ionospheric and thermospheric models. The most direct implementations correlate F$_{10.7}$ with observable atmospheric parameters, such as thermospheric temperature \citep{Jacchia1970} and density \citep{Bowman2008} without directly considering details associated with atmospheric absorption. Models interested in capturing the atmospheric response to solar spectral variability instead parameterize EUV emissions and then simulate the energy deposition into the atmosphere based on its absorption profile and physical state.

One such example is the EUV flux model for aeronomic calculations, EUVAC \citep{Richards1994}. The EUVAC model creates a coarse spectrum covering 5 -- 105 nm in 37 partially overlapping spectral bands \citep{Solomon2005} based on F$_{ave}$ observations by linearly interpolating between two spectra observed on days with significantly different activity levels. In this way, models like EUVAC use a single frequency F$_{10.7}$ measurement to parameterize the entire EUV spectrum.

\begin{figure}[t] 
	\centering
	\includegraphics[trim=0cm 0cm 0cm 0cm, scale=1.0]{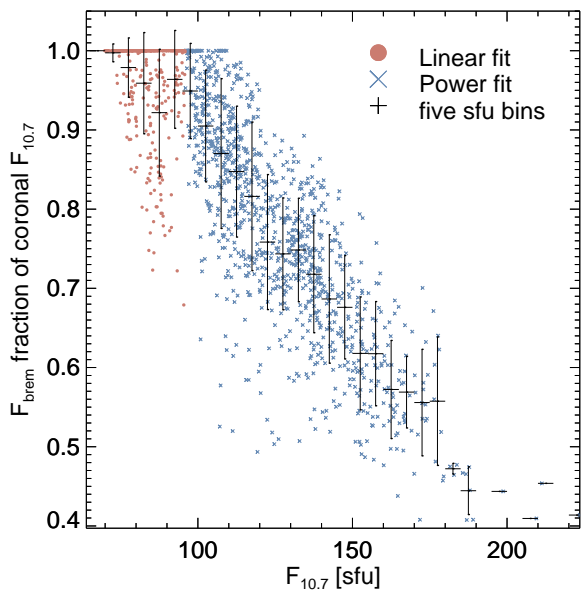}
	\caption{Fraction of coronal F$_{10.7}$ (F$_{brem}$+F$_{gyro}$) associated with the bremsstrahlung emission component. There is a clear trend of decreasing bremsstrahlung fraction with increasing F$_{10.7}$, with F$_{brem}$ accounting for only about 50\% of the coronal contribution when F$_{10.7}$=175. The variation in this bremsstrahlung fraction is well constrained for a given F$_{10.7}$ level.}
	\label{fig:f10_coronal_ex_vchrom}
\end{figure}

The EUVAC model assumes that the observed variability in F$_{10.7}$ reflects variability in coronal EUV emission. The decomposition of F$_{10.7}$ into its various components allows us to test that assumption. Using F$_{brem}$ and F$_{gyro}$ from Figure \ref{fig:component_series_variable_chromosphere} we can examine the fraction of coronal F$_{10.7}$ (F$_{brem}$+F$_{gyro}$) that is attributable to F$_{brem}$ as plotted in Figure \ref{fig:f10_coronal_ex_vchrom}. This illustrates the fact that in the linear-fit regime, the bremsstrahlung emission accounts for nearly all of the coronal F$_{10.7}$. For the power fit, this fraction decreases approximately linearly with increasing activity, dropping below half at the maximum activity observed in this study. The decreasing fraction is due to the highly variable contribution from gyroresonance emission, which necessitates the use of F$_{ave}$ in model applications. However, the deviation in the bremsstrahlung fraction at a given activity level is typically $\pm7\%$, significantly more constrained than the overall variability due to gyroresonance. This suggests the potential for significant improvements when using F$_{fit}$ in place of F$_{ave}$.


\begin{figure*}[t] 
	\centering
	\includegraphics*[trim=0cm 0cm 0cm 0cm, scale=1.0]{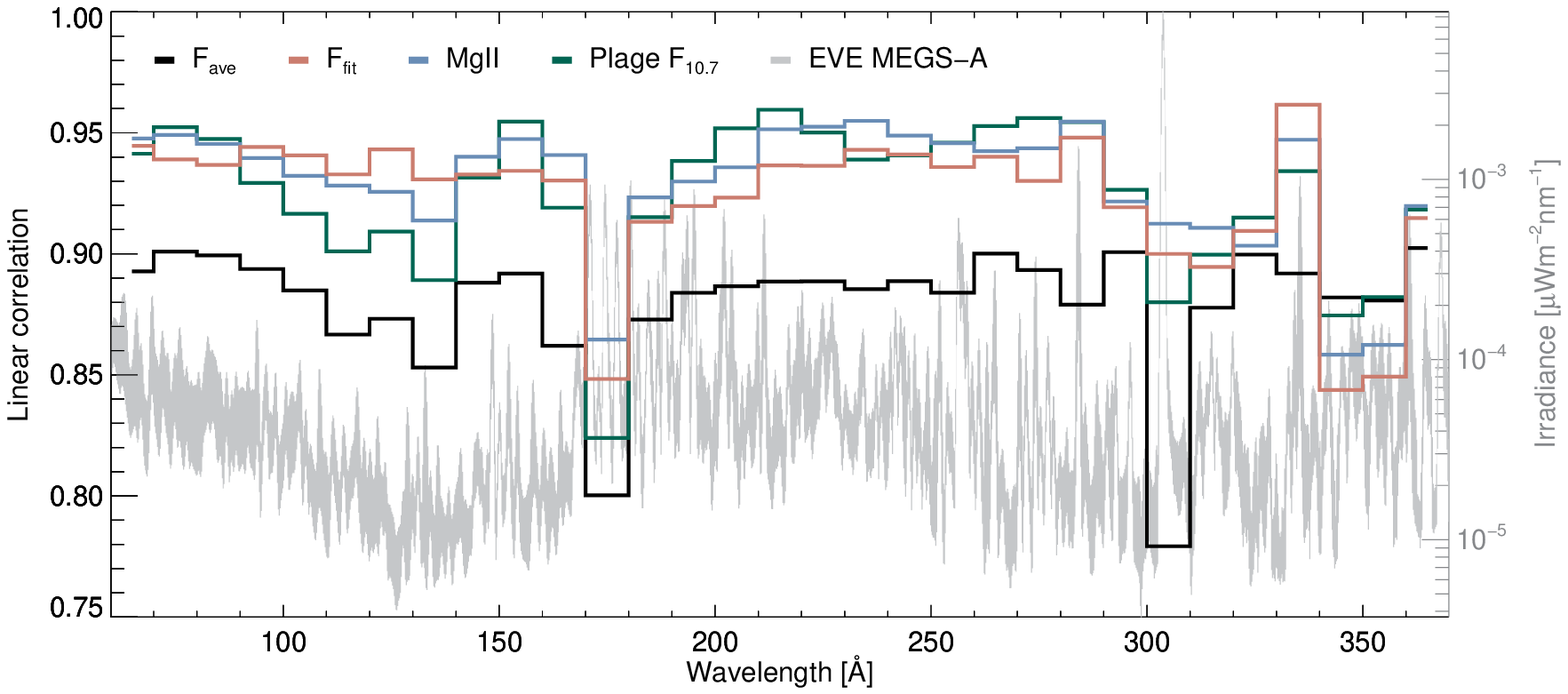}
	\caption{Correlation coefficients between activity proxies and EUV irradiances observed by MEGS-A in one nm bands. The gray region indicates the range of observed MEGS-A daily median spectra. The F$_{fit}$ (red), \ion{Mg}{2} (blue), and plage-field component from \cite{Henney2012a} (green) have similar correlations across this spectral range. Each of these proxies correlate with the observed EUV better than F$_{ave}$ (black) in nearly all spectral bands, including markedly greater correlation in the band containing the \ion{He}{2}, 304 \AA\ line, the single brightest line in the MEGS-A spectral range.}
	\label{fig:f10_spec_correlation_bin}
\end{figure*}                                                                                                                                                                             

The correlation of the MEGS-A spectra with multiple EUV proxies is plotted as a function of wavelength in Figure \ref{fig:f10_spec_correlation_bin}. Across nearly the entire MEGS-A spectral range, F$_{ave}$ has the worst correlation with the EUV observations compared with the other proxies examined in this paper. However, in an absolute sense F$_{ave}$ is still an effective proxy with a Pearson linear correlation coefficient typically greater than 0.85. All of the other tested proxies, F$_{fit}$, the \ion{Mg}{2} index, and the plage component of the photospheric magnetic field proxy from \cite{Henney2012a}, have correlations greater than 0.9 in most spectral bins. Interestingly, the two bins with the worst correlations with F$_{ave}$ (centered at 175 and 305 \AA) are also the bins containing the strongest emission lines, a complex of coronal iron lines and \ion{He}{2} 304 \AA, respectively. These two bins have the smallest relative variation (between minimum and maximum) in the EVE MEGS-A spectra which may be responsible for their slightly reduced linear correlations. The other proxies improve on F$_{ave}$ significantly in the \ion{He}{2} 304 \AA\ bin, the single brightest band in this spectral range containing $\approx$15 -- 30\% of the irradiance observed by MEGS-A, depending on the activity level.



Figure \ref{fig:f10_spec_correlation_bin} demonstrates that F$_{fit}$ generally correlates with the observed EUV better than F$_{ave}$ and as well as \ion{Mg}{2} and the \cite{Henney2012a} index while requiring only a single F$_{10.7}$ measurement. This means it should be possible to improve current atmospheric models and EUV spectral parameterizations simply by using the best-fit bremsstrahlung component of F$_{10.7}$ instead of F$_{ave}$. F$_{fit}$ has the improved correlation of \ion{Mg}{2} while maintaining the long observational history of F$_{10.7}$ and without the risk associated with needing to observe from space. It has the further benefit that it does not require foreknowledge of the future 40 days of F$_{10.7}$ which is required to compute F$_{ave}$, making it potentially even more effective in an operational application. Though the fit was performed using a broad range of solar conditions near minimum to solar maximum, one drawback of F$_{fit}$ is that it is generated with data from a single partial \citep[and fairly weak,][]{Komitov2013, Huang2016} solar cycle and the exact fit parameters may change slightly between solar cycles. However, the efficacy of this fit can be evaluated using current models, and it was performed using a broad range of solar conditions from near minimum to solar maximum.	

\section{Conclusion}
\label{sec:conclusion}
Using four years of DEM results from Paper 1 we investigated the physical emission components of the solar F$_{10.7}$ index. This analysis assumed F$_{10.7}$ is a combination of three independent emission components: optically thick bremsstrahlung from the chromosphere, optically thin bremsstrahlung from the corona, and optically thick gyroresonance from the corona. From the DEMs computed in Paper 1 we directly calculated the coronal bremsstrahlung emission and found that it accounts for $14.2 \pm 2.1$ sfu ($\sim20\%$) of the solar minimum F$_{10.7}$ and more than $40\%$ of the total during more active periods.

We also fit this relationship with a continuously differentiable piecewise function that is linear at low activity levels and a power function at high activity levels. The variable chromosphere was determined by correlating the \ion{Mg}{2} activity proxy with the non-bremsstrahlung F$_{10.7}$ during the linear, low-activity component of the fit. The remaining F$_{10.7}$ is attributed to gyroresonance emission which accounts for almost none of the coronal F$_{10.7}$ in the linear-fit regime and up to more than half during the most active periods of the power-fit regime. This analysis explains the historic disagreement in the literature about the relative contribution of bremsstrahlung and gyroresonance emission in F$_{10.7}$. While bremsstrahlung emission typically contributes the majority of the coronal emission, gyroresonance tends to dominate the rotational modulation due to the discrete nature of active regions.

The piecewise fit to the bremsstrahlung emission as a function of F$_{10.7}$ can be used to define a new activity proxy. We find that this achieves a correlation with observed EUV that is significantly better than F$_{ave}$ and is comparable with using \ion{Mg}{2} or a photospheric magnetic field index. In addition to its improved correlation, this fit proxy has two distinct advantages over the traditional F$_{10.7}$ averaging used to characterize solar EUV variability. First, using the bremsstrahlung trend requires only a single daily F$_{10.7}$ observation. This is preferable if F$_{10.7}$ is used in an operational setting where EUV is monitored daily and in real time because it does not require 40-day foreknowledge like the averaging method. Second, the best-fit trend line provides an uncertainty in the bremsstrahlung prediction. This can be translated into a characteristic range of possible EUV irradiance which allows for the uncertainty in the solar input to be properly accounted for in models.

It is possible that the exact parameterization of the bremsstrahlung component identified here is not the true relationship since it encompasses only the rising phase of a single solar cycle. We might expect a slightly different relationship during the decline of a solar cycle or even minor changes in this relationship between cycles. In addition, the analysis presented does not contain the contributions of a true solar minimum, which would greatly help anchor the linear regime and the calculation of the chromospheric contribution. S.M. White et al. (2019, in preparation) will analyze data from a single EVE sounding rocket calibration flight during solar minimum, which provides an additional constraint on the true solar minimum characteristics of this emission.

There has been considerable concern in recent years that F$_{10.7}$ is insufficient for modern applications \citep{Chen2011a} and should be replaced by other EUV proxies \citep{Tobiska2008, Maruyama2010, Maruyama2011, DudokdeWit2011}, potentially including microwave observations at 30 cm \citep[1 GHz,][]{DudokdeWit2014a}. Our analysis suggests there is still significant value in using F$_{10.7}$ when the physical characteristics of its emission components are considered and the bremsstrahlung component is isolated. This is particularly true in light of its long legacy and familiarity within the community. Even so, this F$_{fit}$ parameterization needs to be tested against observed ionosphere/thermosphere variability to determine if it provides a similar improvement as suggested by its correlation with EUV. In addition, there are situations (e.g., higher cadence input or specific spectral bands that are better modeled with another proxy) that could motivate moving away from F$_{10.7}$ even when it remains effective in its current usage.

\acknowledgements
\textit{Acknowledgements:} Data supplied courtesy of the SDO/EVE consortium. SDO is the first mission to be launched for NASA's Living With a Star (LWS) Program. The F$_{10.7}$ data are provided by the National Research Council of Canada, with the participation of Natural Resources Canada, and support from the Canadian Space Agency. CHIANTI is a collaborative project involving George Mason University, the University of Michigan (USA), and the University of Cambridge (UK). This research has been made possible with funding from AFOSR LRIR 14RV14COR and 17RVCOR416, FA9550-15-1-0014, NSF Career Award \#1255024, PAARE NSF:0849986. S. J. Schonfeld's research was supported by an appointment to the NASA Postdoctoral Program at the Goddard Space Flight Center, administered by Universities Space Research Association under contract with NASA.

\bibliography{library}
\listofchanges
\end{document}